\documentclass[fleqn,usenatbib]{mnras}

\usepackage{newtxtext,newtxmath}

\usepackage[T1]{fontenc}
\usepackage{ae,aecompl}

\usepackage{ulem}

\usepackage{graphicx}	
\usepackage{amsmath}	
\usepackage{amssymb}	

\title[Birth and growth of a solar wind cavity around a comet]{The birth and growth of a solar wind cavity around a comet - Rosetta observations}

\author[E. Behar et al.]{
E. Behar, $^{1,2}$\thanks{E-mail: etienne.behar@irf.se}
H. Nilsson, $^{1,2}$
M. Alho, $^{3}$
C. Goetz, $^{4}$
B. Tsurutani $^{5}$
\\
$^{1}$ Swedish Institute of Space Physics, Kiruna, Sweden.\\
$^{2}$ Lule\aa\ University of Technology, Department of Computer Science, Electrical and Space Engineering, Kiruna, Sweden.\\
$^{3}$ Aalto University, School of Electrical Engineering, Department of Electronics and Nanoengineering, P.O. Box 15500, FI-00076 Aalto, Finland\\
$^{4}$ Technische Universit\"{a}t Braunschweig, Institute for Geophysics and Extraterrestrial Physics, Mendelssohnstra\ss e 3, D-38106 Braunschweig, Germany.\\
$^{5}$ Jet Propulsion Laboratory, California Institute of Technology, 4800 Oak Grove Drive, Pasadena, CA 91109, USA.
}

\date{Accepted 2017 July 20. Received 2017 July 18; in original form 2017 March 28}

\pubyear{2017}

\begin{document}
\label{firstpage}
\pagerange{\pageref{firstpage}--\pageref{lastpage}}
\maketitle

\begin{abstract}

	The Rosetta mission provided detailed observations of the growth of a cavity in the solar wind around comet 67P/CG. As the comet approached the Sun, the plasma of cometary origin grew enough in density and size to present an obstacle to the solar wind. Our results demonstrate how the initial slight perturbations of the solar wind prefigure the formation of a solar wind cavity, with a particular interest placed on the discontinuity (solar wind cavity boundary) passing over the spacecraft. The slowing down and heating of the solar wind can be followed and understood in terms of single particle motion. We propose a simple geometric illustration that accounts for the observations, and shows how a cometary magnetosphere is seeded from the gradual steepening of an initially slight solar wind perturbation. A perspective is given concerning the difference between the diamagnetic cavity and the solar wind cavity.

\end{abstract}

\begin{keywords}
acceleration of particles -- plasmas -- methods: data analysis -- techniques: imaging spectroscopy -- comets: individual: 67P/Churyumov--Gerasimenko
\end{keywords}

\section{Introduction}
	
	Different collisionless plasma regions do not mix easily. When two plasma components meet, structures form, such as collisionless shocks and cavities in stellar and interstellar winds \citep{marcowith2016rpp}. Since the early days of the space age, space missions have made possible the in-situ study of the interaction between the solar wind and various bodies in the solar system. Since then, our understanding of such interactions has gained in precision, with comprehensive studies conducted at planets, moons and small bodies. Comets were no exception in this large scale investigation, with the encounter in 1985 between the International Cometary Explorer (ICE, launched as ISEE-3, see for example \citet{smith1986science}) and comet 21P/Giacobini--Zinner paving the way for subsequent successful cometary missions.
	
	 From a plasma physics perspective, comets distinguish themselves from other bodies of the solar system by three main aspects, namely the eccentricity of their orbit, the composition of their nuclei and their low gravity field. Not gravitationally bound to its source, the cometary atmosphere (coma) grows and shrinks in size and density following the  comets' remoteness from the Sun, which provides the necessary radiation for the sublimation of the various volatiles embedded in their nuclei. From a bare nucleus to a wide and tenuous ionized atmosphere, comets present the most changeful obstacle to the solar wind. In turn, the dynamics of the supersonic solar wind around and through a coma doesn't quite lend itself to comparison with the more familiar planetary picture. In this interaction and during most of the comets' orbital period, the solar wind does not meet a conductive ionosphere or a conductive body: the charge carriers inducing a magnetic response to the solar wind are in fact created (ionised) {\it within} the solar wind, immediately transported away as pick-up ions, and constantly renewed by the ionisation of the neutral coma invisible to the solar wind. Therefore analogies with induced magnetospheres around other unmagnetised bodies are far from trivial.
	
	One unique phenomenon arises in this interaction: the formation of a void in a plasma stream. Unlike previous missions where spacecraft passed far from their target nuclei, at heliocentric distances of around 1 AU, the European Rosetta spacecraft has escorted the comet 67P/Churyumov-Gerasimenko (67P/CG) for two years, along an orbit arc including perihelion, and spanning over heliocentric distances between 1.2 and 4.8 AU. In addition to this large heliocentric distance range, the probe remained closely bound to the nucleus, spending around 80\% of its active mission closer than 200 km to the body center (60\% closer than 100 km). 
	Through Rosetta plasma data, we witness the growth of a solar wind cavity, from initial disturbances of the stream to an established volume devoid of solar wind ions. Later on, as the comet magnetosphere develops, the magnetic field itself is repelled in the vicinity of the nucleus, forming the diamagnetic cavities reported by \citet{goetz2016aa} at 67P/CG, and earlier by \citet{neubauer1986nature} at comet Halley. Rosetta was in fact only the second spacecraft to reach and probe the diamagnetic cavity -- or field-free region --, together with Giotto (with the exception made of the AMPTE mission, in the context of an artificial comet, see for example \citet{haerendel1986nature}). In previous descriptions of the Halley-type cometary plasma environment (see for example the comprehensive review of \citet{neugebauer1990rg}), the diamagnetic cavity surface is defined as the ionopause, or contact surface, and is not distinguishable from the surface of the region free of solar wind ions, referred in this work as the solar wind cavity. One of the seminal models of the interaction between the solar wind and a coma for a high activity comet close to the Sun \citep{biermann1967sp} also describes such a single boundary, separating a region devoid of magnetic field and solar wind ions from the rest of the cometosheath (in which the shocked solar wind ions and cometary ions are mixed). In the environment of 67P/CG, the total plasma density in both types of cavities is still high, and the plasma there is almost purely composed of cometary ions, as shown by \citet{nilsson2017mnras}. Using Rosetta observations, it turns out that at 67P/CG, the solar wind cavity and the diamagnetic cavity are two overlapping but well-separated regions. These differences are shown in the next sections.
	
	In the following study, we look at the solar wind dynamics prefiguring the formation of the solar wind cavity, and depict the discontinuity slowly passing over the spacecraft. After perihelion passage, as the comet draws away from the Sun, the same discontinuity will cross the spacecraft position once more as the cavity shrinks. Contrary to other spacecraft observations obtained during fast passages through boundaries around planets, Rosetta remained close to the nucleus, and plasma structures evolved and moved over the spacecraft, which provide a very high time resolution picture, at the expense of spatial resolution.\\

   The study of \citet{behar2016aa} showed that the interaction changed in character as mass-loading increased, prompting the authors to limit their study to an initial regime of light mass-loading. We extend our study here to also include the medium-to-heavy mass-loading regime. Light mass-loading was already described in \citep{broiles2015aa}, \citep{behar2016grl} and \citep{behar2016aa}. The basic expectations for such a regime are deflection and deceleration of the solar wind, which loses the momentum and the energy gained by the new born cometary ions, themselves accelerated by the convective electric field. But the results from measurements and simulations show that deflection clearly happens, whereas deceleration is very low: the solar wind is interpreted as gyrating in a disturbed magnetic field, with little energy loss \citep{behar2016aa}. Clear deceleration was observed at 1P/Halley \citep{johnstone1986nature, formisano1990aa, neugebauer1990rg}, at 26P/Grigg-Skjellerup \citep{johnstone1993aa}, as well as in the coma of comet 19P/Borrelly by the DeepSpace1 probe \citep{young2004icarus}, and deflection of the solar wind from the comet-Sun line was also reported by \citep{formisano1990aa} at comet 1P/Halley, and at 26P/Grigg-Skjellerup by \citet{johnstone1993aa}. \\
   
    This study is the first to address the regime of heavy mass loading, during which the deflection of the solar wind surpasses 90$^{\circ}$, and when the distribution of the solar wind cannot be considered unidirectional anymore, at the point of observation. \\

\section{Instrument and methods}

The data set used in this study was produced by the Ion Composition Analyzer (ICA), a positive ion spectrometer part of the Rosetta Plasma Consortium (RPC). RPC-ICA \citep{nilsson2007ssr} measures three-dimensional distribution functions of positive ions in the plasma environment of 67P/CG during the entire Rosetta active mission. The instrument detects particles with energy ranging from 10 eV to 40 keV, with a field of view $90^\circ \times 360^\circ$. A complete angular scan is produced in 192 s, with an angular resolution of $5^\circ \times 22.5^\circ$. The instrument mass resolution enables one to distinguish solar wind protons and solar wind alpha particles from each other, and from the rest of the detected particles. Because of technical issues, together with various unsuitable instrument modes that ran during the mission, this study doesn't have a complete time coverage of the 787 days of mission.\\

The solar wind ions were manually selected through the whole set at a daily level, using (energy, mass) selection windows. We refer to \citet{behar2016aa}, where the exact same method is used and illustrated in details. After isolating the two groups of solar wind ions, namely protons $H^+$ and alpha particles $He^{++}$, plasma moments are computed by integrating the measured distributions. Velocity appears in the moment of order 1, the density flux, and is expressed as

\begin{equation}
     \bar{\mathbf{v}}(\mathbf{r})  = \frac{1}{n(\mathbf{r})} \int_{\bf v} \mathbf{v}(\mathbf{r}) f({\bf v}) \ d^3{\bf v} 
\end{equation}

with $\mathbf{r}$ the physical position ({\it i.e.} the spacecraft position) and $\mathbf{v}$ the position in velocity space. $\bar{\mathbf{v}}(\mathbf{r})$ is no more no less than  the center of mass of the same distribution. We refer to ${\bar{\bf v}}$ as bulk velocity.\\

We compute a second value, $\overline{|\mathbf{v}|}$, which directly corresponds to the mean energy of the particles, independent of the direction of their velocity. $\overline{|\mathbf{v}|}$ is defined as
\begin{equation}
   \overline{|\mathbf{v}|}(\mathbf{r})  = \frac{1}{n(\mathbf{r})} \int_{\bf v} |\mathbf{v}(\mathbf{r})| f({\bf v}) \ d^3{\bf v} 
\end{equation}

	The deflection angle from the Sun-comet line (usually simply referred to as deflection) is the angle between the bulk velocity and the anti-sunward direction. It is therefore within $[0^{\circ}, 180^{\circ}]$.\\

	In figure \ref{figAll}, this deflection is given in the two first panels, for solar wind protons and alpha particles. Results are displayed using the following method. Daily time series of proton deflection angles are computed, each value corresponding to a 192 s integration time (full field of view scan). These daily time series are then transformed into histograms of an arbitrary number of bins. One such histogram is given as a vertical set of circles in figure \ref{figAll}: the horizontal position is the date of this histogram, the vertical position is the value of the bin, {\it i.e.} deflection angle, and the radius of the circle gives the number of occurrences for that bin. Occurrences are normalised: the daily sum of occurrences is 1.0 .

	Two different proton speeds are given in the third panel from top, namely the norm of the bulk velocity $|\overline{\mathbf{v}}|$ and the mean speed $\overline{|\mathbf{v}|}$. Each data point corresponds to a daily mean value. The accumulated integration time differs from one data point to the next, due to technical constraints.
	
	The bottom panel gives heliocentric and cometocentric distances.\\

	For each full field of view scan, the integrated number of counts is compared to the count background level (for an equal integration time and an equal Micro Channel Plate detector area), as a ratio. This signal to noise ratio is used to filter all calculated moments, before computing histograms and mean values. A threshold of 10 was used here.
	
	Two periods are indicated with grey background. The first one, early in the mission, corresponds to a time when the Sun was obstructed by the spacecraft from the RPC-ICA field of view. This obstruction, together with the way it was dealt with on-board, results in poor quality moments, as seen on the figure. During the second period, because of technical issues, no data were produced. \\
	
	Integrated ion velocity distributions are given in figure \ref{figDistrib}. The integration time was chosen to be 160 minutes (50 full field of view scans) for each distributions, at the exception of the 6th of August 2014, first day of the active mission (21 scans available only). The distributions are rotated from the instrument reference frame in which they are measured, to a right-handed frame with the x-axis pointing to the Sun and the z-axis direction given by the projection of the proton bulk velocity within the (y, z)-plane. With the assumption that solar wind proton flow remains in a plane while being deflected (assumption corresponding to the light mass-loading case, \citet{behar2016grl}), this frame of reference is an estimation of the Comet-Sun-Electric field frame (CSE). The CSE frame has its x-axis pointing to the Sun, its z-axis along the upstream E-field and the upstream B-field lying in the (x, y)-plane. A similar frame is also used at other bodies, {\it e.g.} the Mars-Sun-Electric field (MSE) frame of reference. 
	
    In figure \ref{figDistrib}, a red dashed circle of constant speed is given for each distribution in the $(v_x, v_z)$ plane. The relevance of this plane is explained in the next section, and the radius of the circle corresponds to the mean energy of the observed protons. This mean energy is taken as the weighted mean of the integrated proton spectrum, and translated into a speed to be displayed in the velocity space.

\section{Results}
	
	The deflection of the solar wind was the first noticeable signature of its interaction with the tenuous ionised coma as activity was slowly increasing ({\it cf.} \citet{broiles2015aa} and \citet{behar2016grl}). The flow remained fairly mono-directional for heliocentric distances greater than 2.75 AU (\citet{behar2016aa}).\\
	
\begin{figure*}
   \begin{center}
  	 \includegraphics[width=\textwidth]{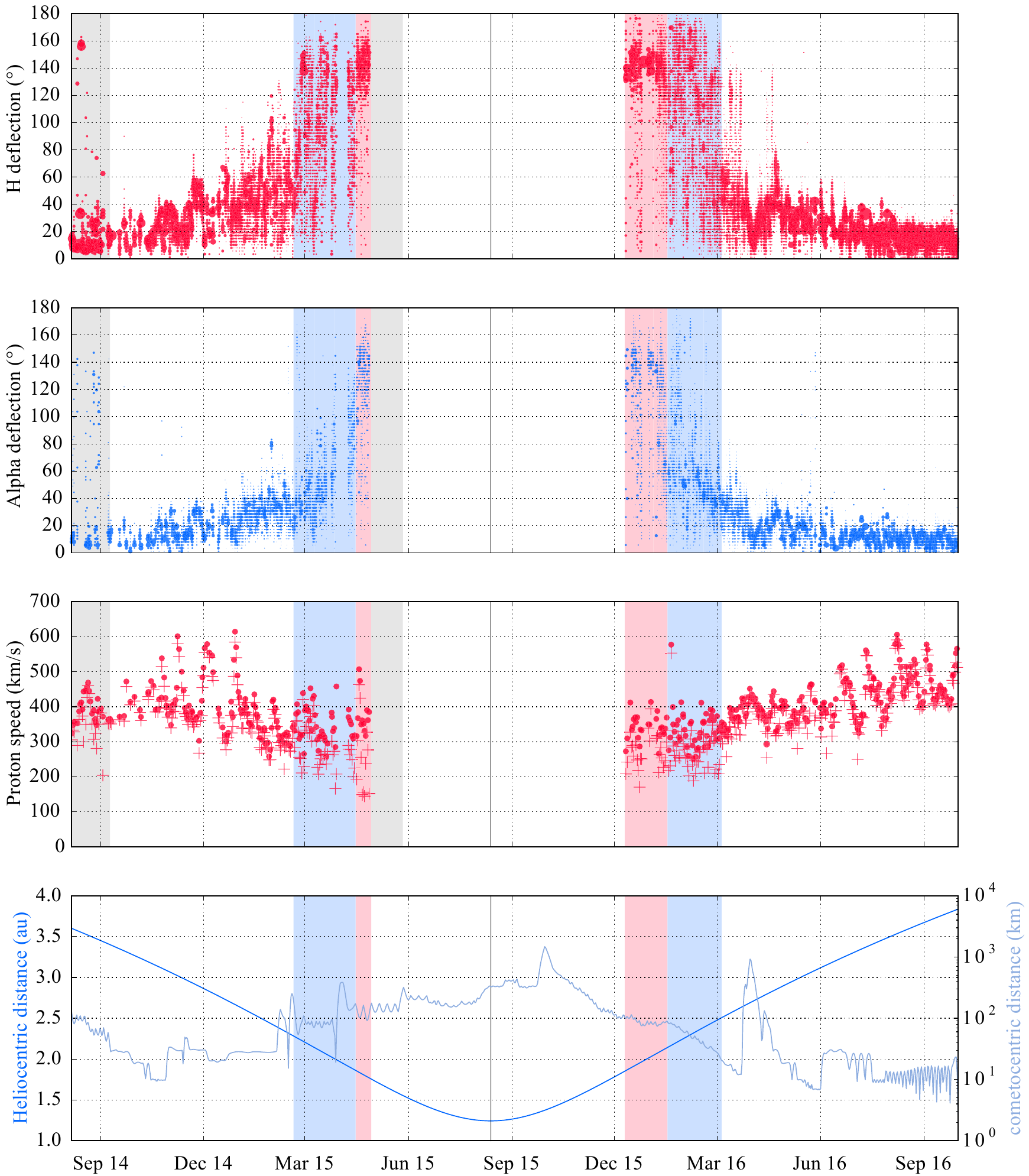}
      \caption{Proton and alpha particle bulk properties through the complete active mission.}
      \label{figAll}
   \end{center}
\end{figure*}

	In figure \ref{figAll}, the spread of this observed deflection is seen to increase later on, closer to the Sun, with a maximum spread reached at distances between 1.9 and 2.5 AU. During this period, highlighted in blue in the same figure, deflection spans from 0$^\circ$ to 180$^\circ$. Note that data are given as daily histograms of plasma moments. Thus, a spread of 180$^\circ$ in one of the displayed histograms is not an indication of a broad shape of the local distribution function, but an indication of an extremely variable distribution, as observed by RPC-ICA. This variability is discussed later in section 4. \\

	During the two periods highlighted with light red background, at distances between 1.64 AU and 2.14 AU, the flow deflection focuses again at $\sim 140^\circ$. During the entire mission, the Sun and the nucleus have been static in the instrument field of view the vast majority of the time, staying in their respective pixel. Therefore, during this refocusing of the deflection, a lack of low deflection values can most likely not be explained as an instrumental effect: we in fact observe a less variable distribution closer to the Sun. \\
	
	In-between these two periods, from 28 April 2015 (1.76 AU), the solar wind is not detected anymore by RPC-ICA, and reappears after perihelion at 1.64 AU on 11 December 2015. We remind the reader that no data are available from 28 April 2015 to 13 May 2015. Protons have been observed during very few events in this time period. These detections are not included in the present results, mostly because of the very low statistics, but such events exist and further case studies should be conducted. One example is given by \citet{edberg2016mnras}, in which authors present a case of a CME hitting the coma, pushing the surface of the solar wind cavity far enough for RPC-ICA to detect protons for a few hours. The spacecraft was then on the way back from a day-side excursion, 800 km away from the nucleus. We also note that alpha particles $He^{++}$ are not observed on the same period (see also \citet{nilsson2017mnras}), which is why we refer to the discussed cavity simply as the solar wind cavity. \\

	The alpha particle deflection in the second panel shows a similar behavior, with noticeably less spread. The value of the deflection is also systematically lower, with the exception of the two red highlighted areas where, just as for protons, alpha particle deflection show a maximum around the same value, $\sim 140^\circ$.

	In-between the two red periods, the spacecraft is not in the solar wind anymore, a solar wind cavity has been created around the nucleus. The most important observation presented in this study might be the following: close to the surface of this cavity, in the terminator plane, the solar wind is flowing almost sunward. \\\\

   The second main expectation for mass-loading is the deceleration of the solar wind, which to first order loses the energy gained by the new born cometary ions, in the comet reference frame. \citet{behar2016aa} reported small deceleration during the regime of light mass-loading, with a maximum difference with the estimated upstream speed of 40 km/s. Because of the lack of direct upstream solar wind parameter measurements, solar wind moments measured at Mars were propagated to 67P/CG, which at that time was lying almost on the same Parker spiral as Mars. Therefore the uncertainty of this deceleration is rather large and hardly quantifiable. Here the period of interest spans over more than two years, thus this technique for upstream speed estimation is not applicable. \\
   
   However, the ion velocity distribution functions can provide some information about deceleration. The choice of frame is motivated by the plasma dynamics described in \citet{behar2016grl}, where data show that in the CSE frame and at low activity, neither the solar wind protons nor the cometary ion flow have a $v_y$-component. When actively rotating the distribution to cancel this $v_y$-component for each measurement (each full field of view), the resulting integrated distribution functions remain very well focused. In figure \ref{figDistrib}, the four first rows correspond to the light-to-medium mass-loading regime, with deflection of the solar wind rising from 0 to 90 degrees. In these distributions, the integrated solar wind signal is perfectly focused, a clear peak is observed. The two last distributions, highlighted with respectively a blue and a red background, were chosen to illustrate the previously mentioned blue and red periods in figure \ref{figAll}. In these two cases, a spread of the distribution is observed along the $v_x$- and $v_z$-axis, the distributions remaining very focused along the $v_y$-axis. In the distribution highlighted with the blue background, a clear partial ring distribution is observed in this plane, and is discussed in the next section.
   
   One thing is important to stress concerning these distributions. Resulting from an integration time of maximum 160 minutes, and with a total number of six, they are not meant to be representative of the complete solar wind dynamics and its details. In other words they don't give the whole story, these distributions are chosen to highlight one of the main and simple aspects of these flow dynamics and their evolution with heliocentric distance. And this aspect -- the solar wind gyrating in the inner coma -- is believed to correspond to the most simple configuration of this interaction.

\section{Discussion}

\subsection{Time variability of the measurements}

    The data show large time variability in the deflection of the solar wind, at the 192 s integration time scale, and especially during the blue highlighted period. In other words, during this period, the direction of the solar wind and its density can change drastically from one scan to another. Fundamentally, only two cases are to be considered: first, the plasma dynamics are not in steady state at this time scale (around 3 minutes) and are very turbulent, and second, the system used for the measurements (the instrument together with the spacecraft) is introducing time dependent effects convolved with the physical signal to be measured. Of course, both cases could be at work at the same time.\\

    A few obvious illustrations for the time variability of the observed system can be listed. In \citet{behar2016grl}, we see for example how the inter-planetary magnetic field can change drastically its direction in less than a minute, and how such changes greatly affect the solar wind direction within the coma. This result indicates that at low activity, the plasma environment around the nucleus is asymmetric, and as the upstream magnetic field changes direction, the spacecraft itself will probe a different region of this asymmetric environment. In such a case, we learn that the system is highly sensitive to upstream conditions.
    
    Another possibility for such a variable solar wind is the intrinsic instability of the interaction. It might be that even for steady upstream conditions, the system itself results in the solar wind never reaching steady state inside the coma. The particle-wave interactions, not taken into account in the present work, could not be ignored for a dedicated study of the short time scale variability of the particle dynamics. Large amplitude, low frequency magnetic waves were detected at 67P/CG, with a period of around 25 s \citep{richter2015ag}, as well as at previously visited comets \citep{brinca1989jgr, tsurutani1995grl}. 
    
    Finally, the structure of the coma and its time variation are obviously to be taken into account. Anisotropic outgassing was observed far from the Sun (more than 3 AU heliocentric distance, {\it cf.} \citet{biver2015aa} and \citet{lee2015aa}) and closer (1.9 AU, \citet{migliorini2016aa}), and a heterogeneous plasma environment is observed in hybrid simulations of the coma-solar wind interaction \citep{koenders2015pss}. \\
    
    On the contrary, not many effects or limitations of the system of measurement can be listed. In fact, the only one which was diagnosed is the limitations of the field of view of the instrument. With a partial coverage of the (azimuth, elevation) space of ($360^{\circ} \times 90^{\circ}$) and a significant portion of this field of view being obstructed by the spacecraft body and solar arrays, distributions will be necessarily only partially observed, or even completely missed, at times. One could argue that this effect could and should be taken into consideration in the analysis, by identifying manually or systematically this type of observations, to then discard them. This could be in fact done for low activity when the deflection and the size of the velocity distribution are still small. But as soon as the distribution functions become much larger and more complex, one simply cannot say whether or not the solar wind signal is completely in the instrument field of view. In that sense, RPC-ICA together with Rosetta is not the ideal system for the observation of the solar wind at higher mass-loading, and this limitation is for now not possible to overcome.\\
 
    On the topic of the observed variability of the solar wind parameters during its interaction with the coma, a multi-instrument study could be conducted to identify properly the physical variations as opposed to instrumental effects. For this study, the influence of the field of view limitations on the results is considered acceptable, with around 53\% of the complete $4\pi$ sr solid angle accessible by RPC-ICA.

\subsection{Gyrating solar wind}

\begin{figure}
   \begin{center}
  	 \includegraphics[width=.5\textwidth]{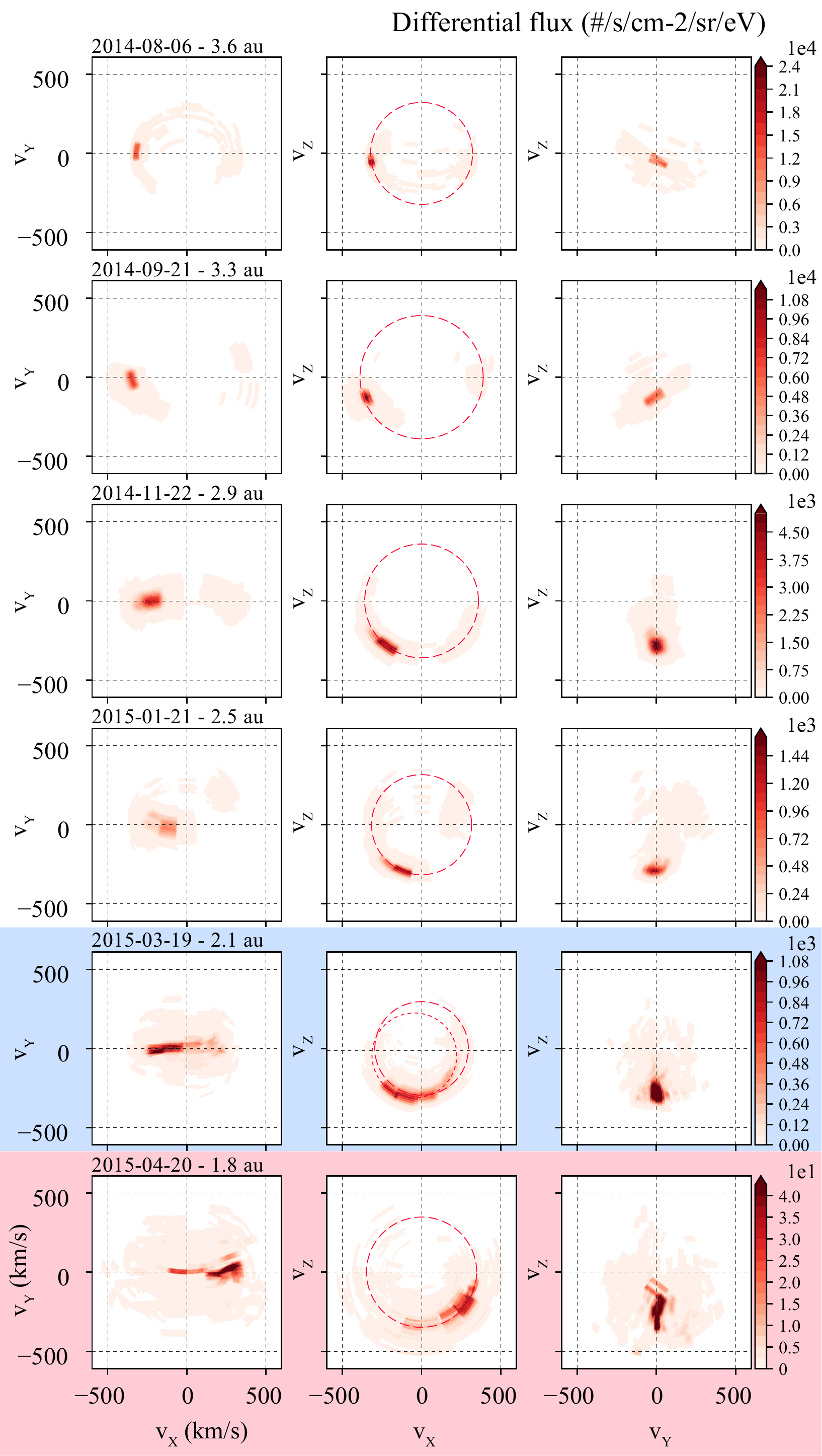}
      \caption{Velocity distributions of solar wind protons during the in-bound leg.}
      \label{figDistrib}
   \end{center}
\end{figure}

   In figure \ref{figAll}, third panel from top, we see that the norm of the bulk velocity is systematically smaller than the mean speed, with a difference between the two growing with activity. As introduced in \citet{behar2016aa}, this can be interpreted as the solar wind gyrating in the coma. In this interpretation, little energy is lost ({\it i.e.} the mean speed is almost conserved from upstream to within the coma), whereas the bulk velocity amplitude is significantly decreased. In figure \ref{figDistrib}, the proton distribution through the in-bound leg remains narrow in terms of speed, as seen in the $(v_x, v_z)$-plane, despite the deflection getting larger with decreasing heliocentric distances. 
   
   The fifth distribution function from top gives us a perfect view of this gyration, with a partial ring distribution in a plane orthogonal to the perpendicular component of the magnetic field. If the loss of energy is small, it should still be expected that the longer a proton interacts with the coma, the larger this loss will be, to the point of being observable by RPC-ICA in this frame. In fact in the same distribution, particles with positive $v_x$-component are heavily deflected, which means they have been interacting with the coma for a longer time than particles with small deflection. Instead of a perfect circle centered on the origin, we should expect a decrease of the radius of this distribution (which directly corresponds to the speed) with an increase of the deflection. Such a trend is clearly visible in this integrated distribution, as in many other distributions during the mission.

   Another interpretation for this characteristic distribution shape can be given, which does not oppose the previous one. In terms of a non-resistive hybrid description, the magnetic field flux tubes are convected at a velocity equal to the electron bulk velocity. In a frame moving with this velocity, the convective electric field disappears and the solar wind particles are seen gyrating without loss of kinetic energy. At low cometary plasma densities, this electron velocity will be dictated by the fast solar wind electrons, much denser than the cometary electrons. At high cometary plasma densities, {\it i.e.} for a dense coma, the opposite happens, and the electron velocity is given mostly by the relatively slow cometary electrons. Therefore the difference between the body centered frame and the frame co-moving with electrons will be small. In this precise case and with this interpretation, we would estimate the electron speed, and therefore the velocity of magnetic field convection, to be around 50 km/s, given by the center of the second smaller circle. This offset has components both along the $v_x$- and the $v_z$-components, indicating that the magnetic field is convected not only anti-sunward along the comet-Sun line but also along the negative $z$-axis, a result also nicely highlighted by \citet{koenders2016mnras}.
    
\subsection{A discontinuity}

\begin{figure*}
   \begin{center}
   \includegraphics[width=\textwidth]{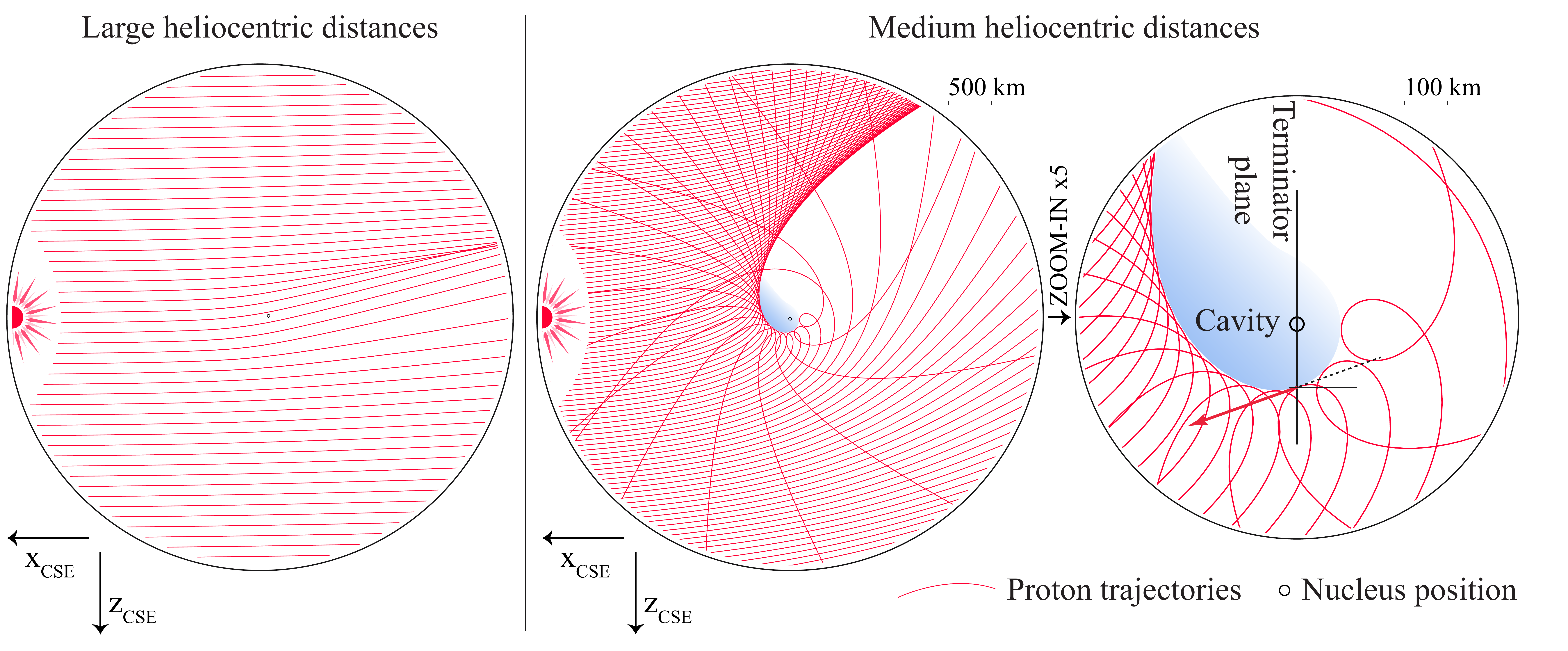}
   \caption{Geometric interpretation of the formation of the solar wind cavity, for the light mass-loading regime and the medium-to-heavy mass-loading regime. \label{interpretation}}
   \end{center}
\end{figure*}

   Through RPC-ICA data, we witness the formation of a solar wind cavity. This cavity can either have a sharp surface, which would be referred to as a discontinuity, or a not well defined surface, in the case of a solar wind density only tending to zero with decreasing distance from the nucleus. The dynamics of the flow for both solar wind species we discuss might tell us something about this surface.\\
  
   In the case of a well defined discontinuity, on the very location of the discontinuity, the flow has to be tangential to the surface. This is a strong constraint on the distribution function: there can be no angular spread outside of the tangent plane. In addition to that, if we consider the solar wind ideally gyrating, the flow has to remain in a single plane normal to the upstream orthogonal component of the magnetic field, at least close to the $(y_{CSE}=0)$ plane. These two simplistic considerations would lead to the expectation of a fairly unidirectional flow at the discontinuity, in the $(x_{CSE}, z_{CSE})$ plane. In fact, the spacecraft can be considered to be very close to this plane, because of its small distance to the nucleus at any time. And we note that just before the solar wind disappearance, and just after its reappearance, the flow deflection is significantly refocused around one value, namely 140$^{\circ}$. The chosen distribution highlighted in red in figure \ref{figDistrib} illustrates the same phenomenon, with a refocusing of the distribution compared to the previous one highlighted in blue.\\ 
   
   Motivated by this peculiar observation, we propose one geometrical illustration bringing these simple and ideal expectations together. We consider the same $(x_{CSE}, z_{CSE})$ plane. We base this illustration on two very simple ideas: first, the solar wind is gyrating in the ionised coma, and second, the closer a proton gets to the nucleus, the shorter its gyroradius is (in other words, the more deflected it gets). This second point is naturally motivated by the fact that ion density, to the first order, is getting larger closer to the nucleus; and the denser the coma gets, the more the solar wind is deflected.
   
    In figure \ref{interpretation}, the Sun is on the left, the nucleus is in the middle of the schematics. Particles are flowing from the left, towards the right. As they get closer to the nucleus, they are more deflected. Every red line is one trajectory, or stream line, and the radius of curvature is a function of the distance to the center. In this precise case, the radius of curvature is proportional to $r^2$. The first panel on the left gives the expectation for a low activity case, and summarises nicely the observations made during this regime: limited deflection, anti-sunward and unidirectional flow. The two other panels give the same type of illustration for a higher activity case. The third panel (right-hand side) is simply a zoom-in of the central one. We see that the solutions of such a configuration can display complete loops along a trajectory, which is the major aspect of this illustration.
    
    We find that an area in which no particle can enter appears around the nucleus: a cavity is formed. In order to give some scale to the picture, we find that with an initial speed of the particles (protons) of 400 km/s and a force deflecting the particles as ${\bf F} = 10^{-11}/r^2$ N (corresponding to the evolution of the radius of curvature discussed above), the extent of the cavity towards the $+z$-direction along the terminator line is of about 150 km, very close to the cometocentric distance at which the spacecraft was when crossing the boundary. What does the discontinuity look like on a terminator plane ({\it i.e.} at the vertical of the nucleus in this illustration, displayed with the solid black line in the lower second panel)? As pointed out above, the flow is in fact tangent to the discontinuity. But even more interesting, at the closest intersection between the terminator line and the discontinuity, this tangent flow (pictured by the red arrow in the third panel) is mostly sunward, with a deflection angle of around 160$^\circ$ . \\
   
   Far from being a self-consistent model, this illustration is nothing more than an attempt to visualise the effect of one simple consideration: what does the simplified general picture look like if the solar wind gyrates in the coma, with a gyro-radius being smaller closer to the nucleus? We see that this results in an asymmetric cavity, generated by purely geometric -- or kinetic -- effects, and that on the terminator plane and at the discontinuity, particles are flowing with a very large deflection angle, almost sunward. Anywhere else in the flow, since different trajectories intersect at any point, the norm of the bulk velocity is smaller than the constant mean speed. Therefore, this illustration accounts for most of the main observations made previously.\\
   
   In the scope of this interpretation, one observation of great interest was made at comet 26P/Grigg-Skjellerup and raised by \citet{jones1997asr}. The authors reported solar wind protons deflected from the comet-Sun line, with a signal consisting of two different beams of similar energy (between 100 and 130 eV), both almost perpendicular to the B-field direction. One peak was reported coming almost from the undisturbed solar wind direction, while the other one was seen much more deflected. Despite a larger deceleration than what was observed at 67P/CG, the geometry of this observation display striking similarities with the present interpretation, in which two well separated beams are seen in most of the area, corresponding to crossing trajectories. A thorough comparison with the observation at 26P/Grigg-Skjellerup would however require a greater modeling effort than presently done.

\subsection{The diamagnetic- and solar wind cavity surfaces separation}

    One of the experimental results brought by Rosetta is the marked difference between the solar wind cavity and the diamagnetic cavity, in contrast with the classical view of the cometary plasma environment given by \citet{biermann1967sp} (Figure 1). Whereas diamagnetic cavity surface crossings have been detected at least 665 times \citep{goetz2016mnras}, we see that the observation of the solar wind cavity consists in one main period of solar wind void, spanning over more than seven months. From the observations, it is clear that during this period the spacecraft was most of the time outside the diamagnetic cavity but inside the solar wind cavity, with a total time spent in the diamagnetic cavity of 42 hours over around 9 months \citep{goetz2016mnras}, compared to only a few proton detections during the similar period. This has a direct implication on the size of these cavities: the solar wind cavity is generally bigger than the diamagnetic cavity. The work of \citet{goetz2016mnras} and the data presented by \citet{edberg2016mnras} give both a feeling about these sizes, respectively 400 km for the diamagnetic cavity (in the terminator plane) and 800 km for the solar wind cavity (in the day-side). These dimensions are only lower limits for cavities observed at different times and different places, close to perihelion.
    
     Using a 2-dimensional bi-ion fluid model, \citet{sauer1994grl} nicely highlighted such a separation between the two boundaries. The authors named the surface of the solar wind cavity the protonopause, an ion composition boundary. On the experimental side, observations of solar wind ions presented in \citet{johnstone1986nature}, \citet{ip1989aj} and \citet{fuselier1988grl} show strong decrease and disappearance of solar wind protons and alpha particles, as well as alpha particle deflection. However, it seems that due to the spacecraft speed and sensors' field-of-views limitations, the precise densities and the precise time of disappearance of these populations are not available for direct comparison with the detected diamagnetic cavity at 1P/Halley. Such a difference of size is also seen in more recent simulation results from \citet{rubin2014aj} (1P/Halley, fluid model) and \citet{koenders2015pss} (67P/CG, hybrid model), both addressing the perihelion case. As demonstrated by \citet{sauer1994grl}, this configuration appears to be similar to the magnetic pile-up region at Venus, where the magnetic field is observed in the absence of solar wind ions ({\it cf} \citet{luhmann1986ssr}, 4.3.1).
    
    A detailed analysis of the possible multiple crossings of the solar wind void surface (at disappearance and reappearance of the solar ions) would require a multi-instrument approach, since the detection relies greatly on the distribution of the solar wind, and therefore on the field-of-view of the instrument and its limitations. \\

\section{Conclusion}
    
    As the nucleus gets closer to the Sun, the two plasma components -- the solar wind and the ionised coma -- tend to separate themselves. The interaction between the solar wind and the comet evolves from an asteroid-like interaction at aphelion, in which the solar wind impacts the surface of the atmosphere-less nucleus, to a Venus-like interaction at perihelion, in which both the solar wind and the diamagnetic cavities are formed. Rosetta allows us to characterise this evolution, and understand how and when the two cavities expand.  \\
    
    A very important aspect of the cavity formation appears to be the gyration of the solar wind: a macroscopic deceleration with limited energy loss, in the comet reference frame. Contrary to the classical fluid and macroscopic treatment of ionopauses, our description of the solar wind interacting with a growing coma does not involve pressure balance, but instead relies on the kinetic behavior of the flow. We do not describe a thermalisation and a decelaration of the solar wind, but instead a macroscopic deceleration together with a conservation of kinetic energy at the single particle scale.\\
    
    At various scales in the universe, plasma stream cavities are to be observed \citep{marcowith2016rpp}. But comets might be the only natural laboratory within our reach for {\it in situ} study where such cavities are forming and vanishing periodically, not because of the change in the incoming flow, but because of the change in the obstacle itself.

\section*{Acknowledgements}

 We wish to thank the referee for her/his detailed comments and constructive indications.\\

The work on RPC-ICA, as well as this PhD project, is supported  by the Swedish National Space Board (SNSB) through grants 108/12, 112/13, 96/15 and 94/11. 

The work on RPC-MAG was financially supported by the German Ministerium f\"ur Wirtschaft und Energie and the Deutsches Zentrum f\"ur Luft- und Raumfahrt under contract 50QP 1401 .

We acknowledge the staff of CDDP and IC for the use of AMDA and the RPC Quicklook database (provided by a collaboration between the Centre de Donn\'{e}es de la Physique des Plasmas (CDPP) supported by CNRS, CNES, Observatoire de Paris and Universit\'{e} Paul Sabatier, Toulouse and Imperial College London, supported by the UK Science and Technology Facilities Council). We are indebted to the whole Rosetta mission team, Science Ground Segment and Rosetta Mission Operation Control for their hard work making this mission possible. 

Portions of this work was done at the Jet Propulsion Laboratory, California Institute of Technology under contract with NASA. \\

The SPICE toolkit from NASA's Navigation and Ancillary Information Facility (NAIF) is heavily used in this study, and made accessible through python by the SpiceyPy wrapper.

\bibliographystyle{mnras}
\bibliography{cometLib}

\bsp	
\label{lastpage}
\end{document}